# Inapproximability for metric embeddings into $\mathbb{R}^d$ [*]

Jiří Matoušek

Department of Applied Mathematics and
Institute of Theoretical Computer Science (ITI)
Charles University, Malostranské nám. 25
118 00 Praha 1, Czech Republic

Anastasios Sidiropoulos

Computer Science and Artificial Intelligence Laboratory
Massachusetts Institute of Technology
32 Vassar St, Room 32G-608, Cambridge, MA, USA

Rev. July/14/2008 JM

**Abstract**

We consider the problem of computing the smallest possible distortion for embedding of a given $n$-point metric space into $\mathbb{R}^d$, where $d$ is *fixed* (and small). For $d = 1$, it was known that approximating the minimum distortion with a factor better than roughly $n^{1/12}$ is NP-hard. From this result we derive inapproximability with factor roughly $n^{1/(22d-10)}$ for every fixed $d \geq 2$, by a conceptually very simple reduction. However, the proof of correctness involves a nontrivial result in geometric topology (whose current proof is based on ideas due to Jussi Väisälä).

For $d \geq 3$, we obtain a stronger inapproximability result by a different reduction: assuming P$\neq$NP, no polynomial-time algorithm can distinguish between spaces embeddable in $\mathbb{R}^d$ with constant distortion from spaces requiring distortion at least $n^{c/d}$, for a constant $c > 0$. The exponent $c/d$ has the correct order of magnitude, since every $n$-point metric space can be embedded in $\mathbb{R}^d$ with distortion $O(n^{2/d} \log^{3/2} n)$ and such an embedding can be constructed in polynomial time by random projection.

For $d = 2$, we give an example of a metric space that requires a large distortion for embedding in $\mathbb{R}^2$, while all not too large subspaces of it embed almost isometrically.

---

# 1 Introduction

Let $\mathbb{X} = (X, \rho_X)$ and $\mathbb{Y} = (Y, \rho_Y)$ be metric spaces and let $f : X \to Y$ be an injective mapping (embedding). The *distortion* of $f$, denoted by $\mathrm{dist}(f)$, is the smallest $D \geq 1$ such that there exists $\alpha > 0$ (a scaling factor) for which $\alpha \rho_X(x,y) \leq \rho_Y(f(x), f(y)) \leq D\alpha \rho_X(x,y)$ for all $x, y \in X$. An embedding with distortion at most $D$ is also called a *D-embedding*. We let $c_\mathbb{Y}(\mathbb{X})$ denote the infimum of all $D \geq 1$ such that $\mathbb{X}$ admits a $D$-embedding into $\mathbb{Y}$.

We will also use the symbol $\Delta(\mathbb{X})$ for the *aspect ratio* of a finite metric space $\mathbb{X}$, which is defined as the largest distance in $\mathbb{X}$ divided by the smallest nonzero distance in $\mathbb{X}$.

Over the past few decades, metric embeddings have resulted in some of the most beautiful and powerful algorithmic techniques, with applications in many areas of computer science [19, 14]. In most of these results, low-distortion embeddings provide a way to simplify the data, without losing too much information.

Here we focus on embeddings of finite metric spaces $\mathbb{X}$ into $\mathbb{R}^d$ with the Euclidean metric $\|.\|$, where $d$ is a *fixed* integer. More precisely, we mainly consider the algorithmic problem of computing or estimating $c_{\mathbb{R}^d}(\mathbb{X})$ for a given $n$-point metric space $\mathbb{X}$. For $d \leq 3$, this problem has an immediate application to visualizing finite metric spaces.

It is known that every $n$-point metric space $\mathbb{X}$ embeds in $\mathbb{R}^d$ with distortion at most $O(n^{2/d} \log^{3/2} n)$ [20]. The proof first embeds $\mathbb{X}$ into a high-dimensional Euclidean space using a well-known result of Bourgain [4], and then projects on a random $d$-dimensional subspace, following a method of Johnson and Lindenstrauss [15]. It provides a randomized polynomial-time algorithm for constructing an embedding with the distortion mentioned above. In particular, it yields an $O(n^{2/d} \log^{3/2} n)$-approximation algorithm for $c_{\mathbb{R}^d}(\mathbb{X})$, and as far as we know, this is the best known approximation algorithm for this problem.

There exist $n$-point metric spaces $\mathbb{X}$ with $c_{\mathbb{R}^d}(\mathbb{X}) = \Omega(n^{1/\lfloor (d+1)/2 \rfloor})$ [20], and thus the above worst-case upper bound cannot be much improved (in particular, for every *even* dimension it is tight up to the logarithmic factor).

Here we will show that, assuming P$\neq$NP, there is no polynomial-time algorithm with approximation ratio *much* better than $n^{2/d}$. Namely, we will prove that $c_{\mathbb{R}^d}(\mathbb{X})$ cannot be approximated with factor smaller than $n^{c/d}$ for a universal constant $c$ (so at least the exponent $c/d$ has the correct order of magnitude as a function of $d$). We now state the results more precisely.

**All dimensions hard...** Bădoiu et al. [6] proved that it is NP-hard to approximate the minimum distortion required to embed a given $n$-point metric space $\mathbb{X}$ into $\mathbb{R}^1$ with factor better than roughly $n^{1/12}$ (see Theorem 3.1 below for a precise formulation). Using their result as a black box, we obtain an analogous hardness result for embeddings in $\mathbb{R}^d$ for every fixed $d \geq 2$:

**Theorem 1.1** *For every fixed $d \geq 2$, and for every fixed $\varepsilon > 0$, it is NP-hard to approximate the minimum distortion required for embedding of a given $n$-point metric space into $\mathbb{R}^d$ within a factor of $\Omega(n^{1/(22d-10)-\varepsilon})$.*

Our derivation of this theorem from the 1-dimensional result is conceptually very simple: Given an $n$-point metric space $\mathbb{X}$, we consider a $(d-1)$-dimensional sphere $S$ in $\mathbb{R}^d$ of radius $R$ much larger than the largest distance in $\mathbb{X}$, and in this $S$ we pick an $\varepsilon$-dense[1] finite set $V$ for a sufficiently small $\varepsilon > 0$. Then we form another metric space $\mathbb{Y} = (Y, \rho_Y)$ as a suitable Cartesian product of $\mathbb{X}$ with $(V, \|.\|)$.

---
[1] A set $V$ in a metric space $(X, \rho_X)$ is called $\varepsilon$-dense if for each $x \in X$ there is $v \in V$ with $\rho_X(x,v) \leq \varepsilon$.

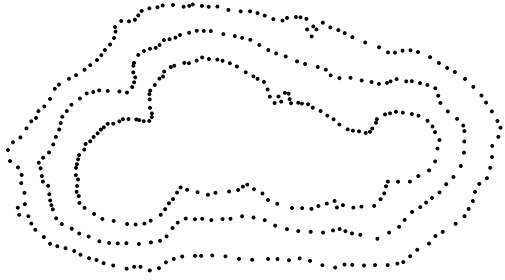

Figure 1: A $D$-embedding of $\mathbb{Y}$ (a schematic illustration for $d = 2$ and $|X| = 3$).

It is easy to show that $c_{\mathbb{R}^d}(\mathbb{Y}) = O(c_{\mathbb{R}^1}(\mathbb{X}))$. The harder part is extracting an $O(D)$-embedding of $\mathbb{X}$ into $\mathbb{R}^1$ from an arbitrary given $D$-embedding $g\colon Y \to \mathbb{R}^d$; see Fig. 1. Intuitively, the image of each copy of $V$ in $\mathbb{Y}$ has to "look like" a deformed sphere, and these "deformed spheres" all have to be nested. Hence they are linearly ordered, and this provides an ordering of the points of $\mathbb{X}$ in a sequence, say $(a_1, a_2, \ldots, a_n)$. Then we define an embedding $f\colon X \to \mathbb{R}^1$ so that $f(a_1) < f(a_2) < \cdots < f(a_n)$, and the difference $f(a_{i+1}) - f(a_i)$ is the distance of the $(i+1)$st "deformed sphere" from the $i$th one.

The claim about the nesting of the "deformed spheres" may seem intuitively obvious, but apparently it is not entirely easy to prove, and for establishing it rigorously we will apply some tools from analysis and from algebraic topology (Section 2); part of the current proof is due to Väisälä [26]. This result and some by-products of the proof can be of independent interest. Then we prove Theorem 1.1 in Section 3 along the lines just indicated.

**... 3 and more dimensions harder?** Theorem 1.1 shows that for embeddability in $\mathbb{R}^d$ it is hard to distinguish bad spaces from even much worse ones. However, for applications of low-distortion embeddings, one is usually most interested in efficiently distinguishing good spaces (embeddable with a constant distortion, say) from bad ones.

For example, Theorem 1.1 leaves open the possibility of a polynomial-time algorithm that, given a metric space $\mathbb{X}$, constructs an embedding of $\mathbb{X}$ into $\mathbb{R}^d$ with distortion bounded by a polynomial in the optimal distortion $c_{\mathbb{R}^d}(\mathbb{X})$. For $d = 1$, there are indeed partial results of this kind for restricted classes of metrics, namely, for weighted trees [6] and for unweighted graphs [8].[2] Thus, at least for these two classes, good and bad embeddability in $\mathbb{R}^1$ can be distinguished efficiently (although in a somewhat weak sense).

Our next result shows that for $d \geq 3$, even this kind of distinguishing good from bad is hard in general:

**Theorem 1.2** *For every fixed $d \geq 3$, it is NP-hard to distinguish between $n$-point metric spaces that embed in $\mathbb{R}^d$ with distortion at most $D_0$, and ones that require distortion at least $n^{c/d}$, where $c > 0$ is a universal constant and $D_0$ is a constant depending on $d$.*

Before proving this result, we first establish a weaker but simpler one in Section 5. The tools developed in this simpler proof also appear in the proof of Theorem 1.2 in Section 6.

---
[2] By a *unweighted graph* we mean a metric space whose point set is the vertex set of a graph $G$ and whose metric is the shortest-path metric of $G$ (where each edge has length 1). Similarly, the metric of a *weighted tree* is the shortest-path metric of some tree, where the edges may have arbitrary nonnegative lengths.

The techniques used in the proof of Theorem 1.2 do not seem to be applicable for the case of embedding into $\mathbb{R}^1$ or $\mathbb{R}^2$. So for $d=1$ or $d=2$, it is still possible that there exists a polynomial-time algorithm that computes an embedding of a given metric space $\mathbb{X}$ into $\mathbb{R}^d$ with distortion at most $c_{\mathbb{R}^d}(\mathbb{X})^{O(1)}$.

**No Menger-type condition for approximate embeddings into the plane.** While we cannot exclude the existence of an efficient algorithm that distinguishes "good spaces from bad ones" for embeddings in $\mathbb{R}^2$, we provide some evidence that obtaining such an algorithm may not be easy, since there is no "local" characterization of good embeddability.

First we recall a well-known lemma of Menger [21], asserting that an $n$-point metric space $\mathbb{X}$ embeds *isometrically* in $\mathbb{R}^d$ if (and only if) every subspace of $\mathbb{X}$ on at most $d+3$ points so embeds. Thus, isometric embeddability into $\mathbb{R}^d$ can be decided locally, by inspecting all $(d+3)$-tuples of points (we should remark that much better algorithms can be obtained by other methods).

In contract to this, we prove the following in Section 7:

**Theorem 1.3** *Let $\varepsilon \in (0,1)$ be given, let $n$ be sufficiently large, and let $1/\sqrt{\varepsilon} \leq k \leq c\sqrt{\varepsilon}n$, where $c$ is a sufficiently small constant. Then there exists an $n$-point metric space $\mathbb{X}$, whose embedding in $\mathbb{R}^2$ requires distortion $\Omega(\sqrt{\varepsilon}\,n/k)$, while every $k$-point subspace can be embedded in $\mathbb{R}^2$ with distortion at most $1+\varepsilon$.*

**Related work.** *Worst-case bounds for embedding in $\mathbb{R}^d$.* Some special classes of metrics are known to embed in $\mathbb{R}^d$ with distortion better than the roughly $n^{2/d}$ upper bound for general metrics: weighted trees with distortion $O(n^{1/(d-1)})$ [12] and ultrametrics with distortion $O(n^{1/d})$ [7]. Unweighted trees [2] and, more generally, unweighted outerplanar graphs [3] embed in $\mathbb{R}^2$ with distortion $O(n^{1/2})$. On the other hand, there exist unweighted planar graphs for which $c_{\mathbb{R}^2}$ is $\Omega(n^{2/3})$ [3].

*Computing low-distortion embeddings in $\mathbb{R}^1$.* We have already mentioned the inapproximability result of Bădoiu et al. [6] concerning embeddings of general metric spaces in $\mathbb{R}^1$. They complemented this result by an $O(n^\beta)$ approximation algorithm for embedding weighted trees in $\mathbb{R}^1$ for some constant $\beta < 1$. For embedding an arbitrary metric space $\mathbb{X}$, they obtained an algorithm with approximation ratio depending on $\Delta(\mathbb{X})$; this was further improved in [5]. For embeddings in $\mathbb{R}^1$ there is also an $O(n^{1/3})$ approximation algorithm for unweighted trees and an $O(n^{1/2})$ approximation algorithm for unweighted graphs [8].

*Embeddings in $\mathbb{R}^2$ and $\mathbb{R}^3$.* In [7] it is shown that it is NP-hard to compute a minimum-distortion embedding into $\mathbb{R}^2$ with the $\ell_\infty$ metric. The same paper gives an $O(n^{1/3})$-approximation algorithm for embedding an ultrametric in $\mathbb{R}^2$, and an $O(\log^{O(1)} \Delta(\mathbb{X}))$-approximation algorithm for this case was given in [23].

*Unlimited dimension and other cases.* Linial et al. [19] observed that an embedding with the smallest possible distortion into $\ell_2$ (or equivalently, into a Euclidean space of an arbitrary dimension) can be computed in polynomial time via semidefinite programming. In contrast, it is well known that deciding *isometric* embeddability in $\ell_1$ is NP-hard (see [11]). Lee et al. [18] obtained an $O(1)$-approximation algorithm for embedding weighted trees into $\ell_p$.

The problem of approximating the minimum distortion embedding has also been studied for the case where we are given two metric spaces $\mathbb{X}, \mathbb{Y}$ of the same cardinality and we want to know $c_\mathbb{Y}(\mathbb{X})$; we refer to [13, 25, 16].

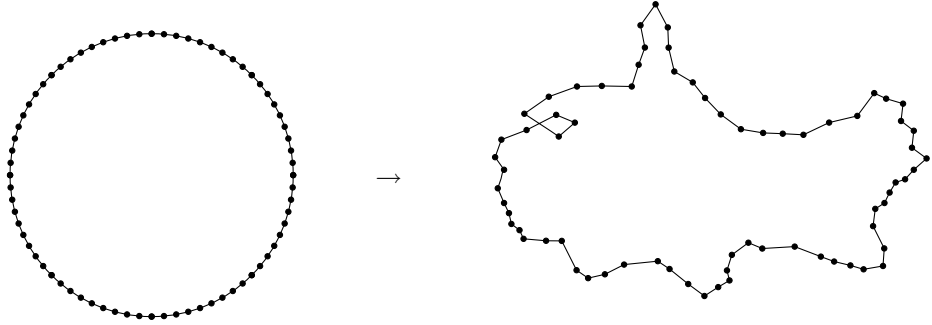

Figure 2: An extension of a $D$-embedding.

*Menger-type questions.* The kind of question which we addressed in Theorem 1.3 for embeddings in $\mathbb{R}^2$ was studied for embeddings in $\ell_1$ by Arora et al. [1], and their results were further strengthened by Charikar et al. [9]. The latter authors proved that if every $k$-point subspace of a an $n$-point metric space $\mathbb{X}$ embeds into $\ell_1$ with distortion $D$, then $\mathbb{X}$ embeds into $\ell_1$ with distortion $O(D\log(n/k))$; moreover, this result is nearly the best possible.

## 2 Deformed spheres and nesting lemmas

As was outlined in the introduction, in the proof of Theorem 1.1 we will be confronted with the following setting: We have a finite set $V$ in a $(d-1)$-dimensional sphere $S$; for the purposes of this section we may assume that $S = S^{d-1}$ is the unit sphere in $\mathbb{R}^d$. We assume that $V$ is $\varepsilon$-dense in $S^{d-1}$, and we are given a $D$-embedding $g: V \to \mathbb{R}^d$. By re-scaling we may assume that $g$ is noncontracting and $D$-Lipschitz.

In order to employ topological reasoning about the image of such $g$, we extend $g$ to a continuous map $\overline{g}: S^{d-1} \to \mathbb{R}^d$ by a suitable interpolation (a tool for doing this will be mentioned in Section 2.2); see Fig. 2. We can make sure that $\overline{g}$ is still $D$-Lipschitz, but generally it won't be noncontracting, and it can even fail to be injective.

However, $\overline{g}$ satisfies the following weaker version of "noncontracting": For all $x, y \in S^{d-1}$ we have $\|\overline{g}(x) - \overline{g}(y)\| \geq \|x - y\| - \delta$, where $\delta = 2D\varepsilon$ (this is easy to check; see Lemma 3.3 below).

The main goal of this section is to show that the image of such $\overline{g}$ behaves, in a suitable sense, as an "approximate sphere". This is expressed in Theorem 2.1 below. For the proof of Theorem 1.2 we will need a technical extension of these results; namely, instead of images of $S^{d-1}$, we need to deal with images of more general shapes, e.g., long tubes and punctured spheres. This is done in Section 4.

### 2.1 Big holes and nested spheres

For a compact set $K \subset \mathbb{R}^d$, let us call a bounded component of $\mathbb{R}^d \setminus K$ a *hole* of $K$.

**Theorem 2.1**

(i) (A big hole exists) *Let $\delta \in [0, \frac{1}{4})$, let $f: S^{d-1} \to \mathbb{R}^d$ be a continuous map that satisfies $\|f(x) - f(y)\| \geq \|x - y\| - \delta$ for all $x, y \in S^{d-1}$, and let $\Sigma := f(S^{d-1})$. Then $\Sigma$ has a hole containing* a ball of radius $\frac{1}{4}$.

(ii) Let $f_1, f_2 : S^{d-1} \to \mathbb{R}^d$ be maps satisfying the condition as in (i) for some $\delta < \frac{1}{4}$, and suppose that, moreover, $\|f_1(x) - f_2(x)\| < \frac{1}{4}$ for all $x \in S^{d-1}$. Then some hole of $\Sigma_1 := f_1(S^{d-1})$ intersects some hole of $\Sigma_2 := f_2(S^{d-1})$.

(iii) (All holes but one are narrow) Let $\delta$, $f$, and $\Sigma$ be as in (i), and let us assume that, moreover, $f$ is $D$-Lipschitz for some $D \geq 1$. Then there is at most one hole of $\Sigma$ containing a ball of radius $4D\delta$.

Part (ii) is what we will need, part (i) can be regarded as a by-product of the proof, and part (iii) we do not need but it comes almost for free and it completes the picture. The main ideas of the proof of (i) and (iii) as given below were found by Väisälä [26] in an answer to a question of the first author, and here they are used with his kind permission (we have independently found another proof, but since it was much less elegant, we reproduce Väisälä's).

**Remark.** The constants in the theorem are generally not optimal. However, in part (i), our method also provides an exact result "in the limit": By modifying the parameters in the proof appropriately (namely, setting $r = r(\delta) = 1 - 2\sqrt{\delta}$ and $\varepsilon = \sqrt{\delta}$ in the proof of Lemma 2.3 below), we obtain the existence of a hole containing a ball of radius $r(\delta)$ that tends to 1 as $\delta \to 0$.

**Remark.** In part (iii), for dimension $d \geq 3$ we cannot claim that all holes of $\Sigma$ but one have small diameter, since there can be many thin but very long holes. For $d = 2$ all holes but one must have a small diameter [26].

Here is the result we need for the proof of Theorem 1.1:

**Corollary 2.2 (Nesting lemma)** *Let $\delta < \frac{1}{4}$ and let $f_1, f_2, \ldots, f_n : S^{d-1} \to \mathbb{R}^d$ be continuous maps satisfying*

- $\|f_i(x) - f_i(y)\| \geq \|x - y\| - \delta$ *for all $x, y \in S^{d-1}$ and all $i$,*
- $\|f_i(x) - f_j(x)\| \leq \frac{1}{4}$ *for all $i, j$ and all $x \in S^{d-1}$, and*
- $\Sigma_i \cap \Sigma_j = \emptyset$ *whenever $i \neq j$, where $\Sigma_i = f_i(S^{d-1})$.*

*Let $U_i$ denote the unbounded component of $\mathbb{R}^d \setminus \Sigma_i$, and let us define a relation $\preceq$ on $[n]$ by setting $i \preceq j$ if $U_j \subseteq U_i$. Then $\preceq$ is a linear ordering on $[n]$.*

**Proof.** It is clear that $\preceq$ is a partial ordering, so it suffices to check that for every $i \neq j$ we have either $U_i \subset U_j$ or $U_j \subset U_i$.

Theorem 2.1(ii) shows that some hole of $\Sigma_i$ intersects some hole of $\Sigma_j$. Since $\Sigma_i \cap \Sigma_j = \emptyset$ and $\Sigma_i$, $\Sigma_j$ are path-connected, $\Sigma_i$ and all of its holes are contained in some component of $\mathbb{R}^d \setminus \Sigma_j$, and vice versa, and the nesting lemma follows. □

## 2.2 A lemma on approximate inverse

The first main step in the proof of Theorem 2.1 is the next lemma, which says that $f$ has an "approximate inverse" mapping $h$ that extends to some neighborhood of $\Sigma$.

**Lemma 2.3** *Let $f$, $\Sigma$, and $\delta < \frac{1}{4}$ be as in Theorem 2.1(i), and let $\Omega_r$ denote the closed $r$-neighborhood of $\Sigma$ in $\mathbb{R}^d$. Then there is a continuous map $h\colon \Omega_{1/4} \to S^{d-1}$ such that $\|h(f(x)) - x\| \le 8\delta$ for all $x \in S^{d-1}$, and (consequently) the composition $hf\colon S^{d-1} \to S^{d-1}$ is homotopic[3] to the identity map $\mathrm{id}_{S^{d-1}}$.*

In the proof we will use a basic result about Lipschitz maps: the Kirszbraun theorem [17], which asserts that every Lipschitz mapping from a subset of a Hilbert space $H_1$ into a Hilbert space $H_2$ can be extended to a Lipschitz map $H_1 \to H_2$, with the same Lipschitz constant.

**Proof of Lemma 2.3.** Let us put $\varepsilon := \delta$ and $r := \frac{1}{4}$. Let $N \subset \Sigma$ be an $\varepsilon$-net[4] in $\Sigma$. We choose a mapping $g\colon N \to S^{d-1}$ with $fg = \mathrm{id}_N$; in other words, for every $y \in N$ we arbitrarily choose $g(y) \in f^{-1}(y)$.

We claim that $g$ is 2-Lipschitz. Indeed, if $y, y' \in N$ are distinct and $x = g(y)$, $x' = g(y')$, then the condition on $f$ gives $\|x - x'\| \le \|f(x) - f(x')\| + \delta = \|y - y'\| + \delta < (1 + \delta/\varepsilon)\|y - y'\| = 2\|y - y'\|$ since $\|y - y'\| > \varepsilon$.

Next, using the Kirszbraun theorem mentioned above, we extend $g$ to a 2-Lipschitz map $\overline{g}\colon \Omega_r \to \mathbb{R}^d$. We check that $0$ is not in the image of $\overline{g}$; indeed, if we had $\overline{g}(y) = 0$ for some $y \in \Omega_r$, we could find a point $z \in N$ at distance at most $r + \varepsilon$ from $y$, hence $\|\overline{g}(z) - \overline{g}(y)\| \le 2(r + \varepsilon) < 1$ (using $r = \frac{1}{4}$, $\varepsilon = \delta < \frac{1}{4}$), but $\overline{g}(y) = 0$ while $\|\overline{g}(z)\| = 1$ since $\overline{g}(z) = g(z) \in S^{d-1}$.

We can now define the desired $h\colon \Omega \to S^{d-1}$ as in the lemma, by $h(y) = \overline{g}(y)/\|\overline{g}(y)\|$.

Given $x \in S^{d-1}$, we pick $z \in N$ at most $\varepsilon$ away from $f(x)$, and we calculate $\|\overline{g}(f(x)) - x\| \le \|\overline{g}(f(x)) - g(z)\| + \|g(z) - x\| \le 2\|z - f(x)\| + \|z - f(x)\| + \delta \le 3\varepsilon + \delta = 4\delta$. Since $\|\overline{g}(f(x)) - h(f(x))\| = 1 - \|\overline{g}(f(x))\| \le \|x - \overline{g}(f(x))\|$, we obtain $\|h(f(x)) - x\| \le 2\|\overline{g}(f(x)) - x\| \le 8\delta$ as claimed.

For $\delta < \frac{1}{4}$, this implies that $h(f(x)) \ne -x$ for all $x \in S^{d-1}$, and consequently, $hf \sim \mathrm{id}_{S^{d-1}}$ (this is a standard and easy fact in topology; if $x$ and $h(f(x))$ are not antipodal, they are connected by a unique shortest arc, and the homotopy moves along this arc). The lemma is proved. $\square$

## 2.3 The Alexander duality

In the subsequent proof of Theorem 2.1, we will use cohomology groups. We will not need their definition, only few very simple properties, which we will explicitly state, plus one slightly deeper result of algebraic topology. These can be taken as purely formal rules, which we will apply in the proof. We consider $(d-1)$-dimensional cohomology, since it is closely related to the number of holes.

Each compact set $X \subset \mathbb{R}^d$ is assigned the $(d-1)$-dimensional Čech (or equivalently, Alexander–Spanier) cohomology group[5] $\check{H}^{d-1}(X)$; for definiteness we consider integer coefficients, although the coefficient ring doesn't matter in our considerations. This $\check{H}^{d-1}(X)$ is an Abelian group, and if it is finitely generated, then it is isomorphic to $\mathbb{Z}^b$ for an integer $b \ge 0$, called the *rank* of $\check{H}^{d-1}(X)$.

---

[3] We recall that two continuous maps $f, g\colon X \to Y$ of topological spaces are *homotopic*, in symbols $f \sim g$, if there exists a continuous map $F\colon X \times [0,1] \to Y$ such that $F(x, 0) = f(x)$ and $F(x, 1) = g(x)$ for all $x \in X$.

[4] We recall that a subset $N \subseteq M$ in a metric space $(M, \rho_M)$ is an $\varepsilon$-*net* if every two points of $N$ have distance greater than $\varepsilon$ and $N$ is inclusion-maximal with respect to this property; that is, every point of $M$ is at most $\varepsilon$ far from some point of $N$.

[5] We need Čech cohomology so that our considerations are valid even for $X$ with various local pathologies. In our application of Theorem 2.1 we can assume that the mapping $f$ is "nice", e.g., that its image $\Sigma$ is the union of finitely many simplices, and then we could work with the perhaps more familiar singular or simplicial cohomology.

A very rough intuition is that the elements of $\check{H}^{d-1}(X)$ correspond to (equivalence classes of) $(d-1)$-dimensional "surfaces" inside $X$, with nonzero elements corresponding to "surfaces" that "enclose" one or more of the holes of $X$. (This is really closer to the idea of homology, rather than cohomology, but hopefully it is not totally misleading for our purposes.)

A special case of the *Alexander duality*, which we will state precisely in the proof of Lemma 2.4 below, tells us that the rank of $\check{H}^{d-1}(X)$ equals the number of holes of $X$. For example, $S^{d-1}$ encloses a single hole, and we have $\check{H}^{d-1}(S^{d-1}) \cong \mathbb{Z}$.

A continuous map $f\colon X \to Y$ of compact sets induces a group homomorphism $f^*\colon \check{H}^{d-1}(Y) \to \check{H}^{d-1}(X)$; we should stress that $f^*$ goes in opposite direction compared to $f$. For the composition of maps we then have $(fg)^* = g^*f^*$ (the last two properties are usually expressed by saying that cohomology is a contravariant functor). Moreover, if $f_1, f_2\colon X \to Y$ are homotopic maps, then $f_1^* = f_2^*$.

The following lemma encapsulates what we will need from the Alexander duality.

**Lemma 2.4 ($(d-1)$-dimensional cohomology and holes)**

(i) *Let $d \geq 2$, let $X \subseteq Y$ be compact sets in $\mathbb{R}^d$, let $j\colon X \to Y$ denote the inclusion map, and let $j^*\colon \check{H}^{d-1}(Y) \to \check{H}^{d-1}(X)$ be the induced homomorphism in cohomology. Then the number of holes of $X$ that contain at least one hole of $Y$ equals the rank of the image $\operatorname{Im} j^*$.*

(ii) *Let $d \geq 2$, let $X_1, X_2, Y$ be compact sets in $\mathbb{R}^d$, $X_1 \subseteq Y$, $X_2 \subseteq Y$, let $j_1, j_2$ be the inclusion maps and $j_1^*, j_2^*$ the induced homomorphisms in cohomology. Suppose that $\operatorname{Ker}(j_1^*) \cup \operatorname{Ker}(j_2^*)$ does not generate all of $\check{H}^{d-1}(Y)$. Then there is a hole of $Y$ contained both in a hole of $X_1$ and in a hole of $X_2$.*

**Proof.** The usual (modern) statement of Alexander duality, for the particular dimensions we are interested in, tells us that for every compact $X \subset \mathbb{R}^d$, there is an isomorphism

$$\alpha_X\colon \check{H}^{d-1}(X) \to \tilde{H}_0(\mathbb{R}^d \setminus X).$$

Here $\tilde{H}_0(.)$ is the 0-dimensional *reduced singular homology* group, which in our case can be concretely represented as follows: Each element $\gamma \in \tilde{H}_0(\mathbb{R}^d \setminus X)$ can be regarded as a function that assigns to every hole of $X$ an integer number (and with only finitely many nonzero values, in case that there are infinitely many holes). The group operation is componentwise addition; thus, $\tilde{H}_0(\mathbb{R}^d \setminus X)$ is a free Abelian group whose rank is the number of holes.

Now let $X \subseteq Y$, let $j\colon X \to Y$ be the inclusion map, and let $i\colon \mathbb{R}^d \setminus Y \to \mathbb{R}^d \setminus X$ be the inclusion map of the complements. We need a property of the Alexander duality called *naturality with respect to inclusion*,[6] which means that

$$\alpha_X j^* = i_* \alpha_Y, \tag{1}$$

where $i_*\colon \tilde{H}_0(\mathbb{R}^d \setminus Y) \to \tilde{H}_0(\mathbb{R}^d \setminus X)$ is the homomorphism in homology induced by the inclusion $i$.

If we represent $\tilde{H}_0(\mathbb{R}^d \setminus Y)$ and $\tilde{H}_0(\mathbb{R}^d \setminus X)$ as above, $i_*$ acts as follows: Given an integer function $\gamma$ on the holes of $Y$, the value of the function $i_*(\gamma)$ on a given hole $U$ of $X$ is the sum of $\gamma(V)$ over all holes $V$ of $Y$ contained in $U$. Thus, the rank of $\operatorname{Im} i_*$ is the number of holes of $X$ that contain some hole of $Y$, and since by (1), $\operatorname{Im} i_*$ and $\operatorname{Im} j^*$ are isomorphic, part (i) of the lemma follows.

---

[6]See Munkres [22] §72–74; the naturality is explicitly stated there only for polyhedral sets, since it is needed for the proof of the general case, but it follows for the general case as well by the limiting process used in the proof.

For part (ii), from the assumption and (1) we get that $\mathrm{Ker}(i_{1*}) \cup \mathrm{Ker}(i_{2*})$ do not generate all of $\tilde{H}_0(\mathbb{R}^d \setminus Y)$. Now $\mathrm{Ker}(i_{1*})$ contains all functions that are nonzero only on the holes of $Y$ not contained in any hole of $X_1$, and similarly for $\mathrm{Ker}(i_{2*})$. There is an element (function) $\gamma \in \tilde{H}_0(\mathbb{R}^d \setminus Y)$ that is not a linear combination of such functions, and such a $\gamma$ has to be nonzero on a hole that is contained both in a hole of $X_1$ and in a hole of $X_2$. $\square$

**Proof of Theorem 2.1.** Let us consider the map $f$ as in part (i), and $\Omega_r$ and $h$ as in Lemma 2.3. Let $j: \Sigma \to \Omega_{1/4}$ denote the inclusion map. The composed map $hf = hjf: S^{d-1} \to S^{d-1}$ is homotopic to the identity, and so the induced map $f^* j^* h^*: \check{H}^{d-1}(S^{d-1}) \to \check{H}^{d-1}(S^{d-1})$ in cohomology is the identity as well. Since $\check{H}^{d-1}(S^{d-1}) \neq 0$, the homomorphism $j^*: \check{H}^{d-1}(\Omega_{1/4}) \to \check{H}^{d-1}(\Sigma)$ cannot be zero. By Lemma 2.4(i) this means that there is a hole of $\Sigma$ that contains a hole of $\Omega_{1/4}$, and such a hole of $\Sigma$ contains a $\frac{1}{4}$-ball.

In part (ii), let $\Omega$ be the $\frac{1}{4}$-neighborhood of $\Sigma_1$, let $j_1: \Sigma_1 \to \Omega$ be the inclusion map, and let $h_1: \Omega \to S^{d-1}$ be as in the proof of (i), i.e., with $h_1 f_1 \sim \mathrm{id}_{S^{d-1}}$. By the assumption $\Sigma_2 \subseteq \Omega$ as well (with inclusion map $j_2$), and $f_1$ and $f_2$ are homotopic as maps $S^{d-1} \to \Omega$, since the segment $f_1(x) f_2(x)$ is contained in $\Omega$ for every $x \in S^{d-1}$. So the homomorphisms $f_1^* j_1^*$ and $f_2^* j_2^*$ in cohomology are equal, and also nonzero, since $f_1^* j_1^* h_1^*$ is the identity in $\check{H}^{d-1}(S^{d-1})$.

The kernels of $j_1^*$ and $j_2^*$ are both contained in $\mathrm{Ker}(f_1^* j_1^*) = \mathrm{Ker}(f_2^* j_2^*)$, and the latter is not all of $\check{H}^{d-1}(\Omega)$. Thus, Lemma 2.4(ii) implies that there is a hole of $\Omega$ contained both in a hole of $\Sigma_1$ and in a hole of $\Sigma_2$, and part (ii) is proved.

For part (iii), let us consider the map $h$ as in Lemma 2.3, but restricted to the domain $\Sigma$. This time we set $r = 4D\delta$, and we let $j: \Sigma \to \Omega_r$ be the inclusion map. In part (i), we considered the composition $hf$; here we look at $fh$. This is a map $\Sigma \to \Sigma$, but we regard it as a map $\Sigma \to \Omega_r$, and we want to show that it is homotopic to $j$.

To this end, it suffices to check that $\|f(h(y)) - y\| \leq 2r$ for all $y \in \Sigma$, since then the segment connecting $f(h(y))$ to $y$ lies completely in $\Omega_r$ and it defines the desired homotopy.

Choosing $x \in S^{d-1}$ with $y = f(x)$, Lemma 2.3 gives $\|h(y) - x\| \leq 8\delta$, and thus $\|f(h(y)) - y\| \leq 8D\delta = 2r$ as needed.

By the homotopy just established, we get that $h^* f^* = j^*$ as homomorphisms $\check{H}^{d-1}(\Omega_r) \to \check{H}^{d-1}(\Sigma)$. But since $\mathrm{Im}\, f^* \subseteq H^{d-1}(S^{d-1})$, which has rank 1, $\mathrm{Im}\, j^*$ has rank at most 1, and thus there is at most one hole of $\Sigma$ that contains a hole of $\Omega_r$. $\square$

## 3 Hardness for $\mathbb{R}^1$ implies hardness for $\mathbb{R}^d$

As was mentioned in the introduction, we derive Theorem 1.1 from the result of Bǎdoiu et al. [6] on inapproximability for embeddings into $\mathbb{R}^1$. By inspecting the full version of that paper (available on-line), one can check that their proof yields the following:

**Theorem 3.1 (Bǎdoiu et al. [6])** *Assuming $P \neq NP$, there is no polynomial-time algorithm with the following properties:*

- *The input of the algorithm is an $n$-point metric space $\mathbb{X}$ with $\Delta(\mathbb{X}) = O(n)$.*

- *If $\mathbb{X}$ admits an $O(n^{4/12})$-embedding into $\mathbb{R}^1$, the algorithm answers YES.*

- *If $\mathbb{X}$ is not embeddable in $\mathbb{R}^1$ with distortion smaller than $\Omega(n^{5/12-\varepsilon})$, the algorithm answers NO.*

This theorem together with the next proposition imply Theorem 1.1 by a simple calculation.

**Proposition 3.2** *Let $\mathbb{X} = (X, \rho_X)$ be an n-point metric space, let $d \geq 2$ be a fixed integer, and let $D_{\max} \geq 1$ be a parameter (specifying the maximum distortions we want to consider). There exists a metric space $\mathbb{Y} = (Y, \rho_Y)$, $|Y| = O(nD_{\max}^{2(d-1)}\Delta(\mathbb{X})^{d-1})$, which can be constructed in time polynomial in $n$, $\Delta(\mathbb{X})$, and $D_{\max}$ (the implicit constants depending on d), with the following properties:*

(i) *If $\mathbb{X}$ can be D-embedded in $\mathbb{R}^1$ for some $D \geq 1$, then $\mathbb{Y}$ can be $(1.1D)$-embedded[7] in $\mathbb{R}^d$.*

(ii) *Given a D-embedding of $\mathbb{Y}$ in $\mathbb{R}^d$ for some $D$, $1 \leq D \leq D_{\max}$, one can construct a $1.1D$-embedding of $\mathbb{X}$ in $\mathbb{R}^1$ in polynomial time.*

**The construction.** We follow the sketch given after Theorem 1.1. Let us assume that the smallest distance in $\mathbb{X}$ is 1 and the largest one is $\Delta$. We let $C = C(d)$ be a sufficiently large constant, we set $R := CD_{\max}\Delta$, and we let $S$ be the $(d-1)$-dimensional sphere in $\mathbb{R}^d$ centered at 0 of radius $R$. We set $\varepsilon := \frac{1}{CD_{\max}}$, and we choose $V$ as an $\varepsilon$-dense subset of $S$ (that is, each point of $S$ has distance at most $\varepsilon$ to some point of $V$); as is well known, we can assume that $V$ has size $O((R/\varepsilon)^{d-1})$ and is computable in time polynomial in $R/\varepsilon$.

Then we let $\mathbb{Y} = (Y, \rho_Y) := \mathbb{X} \times_{L_2} V$; that is, we set $Y := X \times V$, and we define the metric $\rho_Y$ by
$$\rho_Y((a,v),(a',v')) = \sqrt{\rho_X(a,a')^2 + \|v-v'\|^2}.$$

**Proof of Proposition 3.2(i).** Let $f: X \to \mathbb{R}^1$ be a D-embedding; we assume that it is non-contracting and D-Lipschitz, and satisfies $\min f(X) = 0$. We define $g: Y \to \mathbb{R}^d$ by $g(a,v) := (R + f(a))v$, so that the image of $(a,v)$ lies on the sphere of radius $R + f(a)$ concentric with $S$. This embedding is illustrated in Fig. 3 (the picture is not realistic, though, since the radii should be larger and the points denser than shown).

Bounding the distortion of $g$ can be divided into two steps. Formally we "factor" $g$ through an auxiliary metric space $Z := f(X) \times_{L_2} V$ (constructed from $f(X)$ with the metric of $\mathbb{R}^1$ in the same way as $\mathbb{Y}$ was constructed from $\mathbb{X}$). So we define $g_1: Y \to Z$ by $g_1(a,v) = (f(a),v)$ and $g_2: Z \to \mathbb{R}^d$ by $g_2(f(a),v) = g(a,v)$. Now $g = g_2g_1$, so $\text{dist}(g) \leq \text{dist}(g_2)\text{dist}(g_1)$. It is almost obvious (and easy to check) that $\text{dist}(g_1) \leq \text{dist}(f) = D$. Checking $\text{dist}(g_2) \leq 1.1$ (for $C$ sufficiently large) is a simple exercise; geometrically, it amounts to bounding the metric difference between a narrow annulus in $\mathbb{R}^d$ and a narrow cylindric band in $\mathbb{R}^{d+1}$, and we omit it. □

For part (ii), let $g: Y \to \mathbb{R}^d$ be a D-embedding; for convenience, we assume that it is noncontracting and D-Lipschitz. For each $a \in X$ we consider the "slice" of $g$, i.e., the mapping $g_a: V \to \mathbb{R}^d$ given by $g_a(v) = g(a,v)$.

Next, we extend each $g_a$ to a D-Lipschitz map $\overline{g}_a: S \to \mathbb{R}^d$ using the Kirszbraun theorem (mentioned after Lemma 2.3).[8] Let $\Sigma_a := \overline{g}_a(S)$ be the image of $\overline{g}_a$.

---

[7] If needed, we could replace 1.1 by any other constant greater than 1, with appropriate adjustments in other constants. We use 1.1 so that we need not introduce an extra parameter.

[8] Another way of extending the $g_a$ is to assume that $V$ is a vertex set of some fine enough triangulation of $S^{d-1}$, and extend affinely on each simplex of the triangulation. In this way we have more control about the local properties of the image (which is piecewise linear), but we need to worry about the existence of a suitable triangulation.

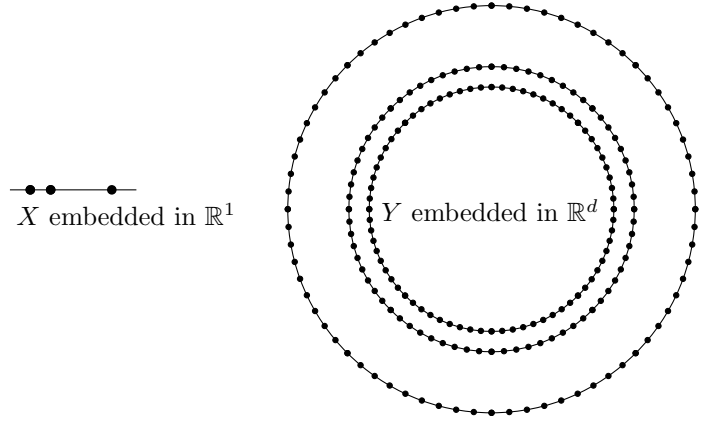

Figure 3: Embedding of $Y$ from an embedding of $X$.

Now we want to use the nesting lemma (Corollary 2.2) to show that the $\Sigma_a$ have to be nested. More precisely, we want to check that by scaling both the domain and range of each $\overline{g}_a$ by $\frac{1}{R}$, we obtain maps $S^{d-1} \to \mathbb{R}^d$ as in Corollary 2.2. This is done using the next two lemmas.

**Lemma 3.3** *For all $x, y \in S$ we have $\|\overline{g}_a(x) - \overline{g}_a(y)\| \geq \|x - y\| - \frac{3}{C}$.*

**Proof.** (Routine.) We find $u \in V$ with $\|x - u\| \leq \varepsilon$, and similarly $v \in V$ for $y$. Then $\|\overline{g}_a(x) - \overline{g}_a(y)\| \geq \|g_a(u) - g_a(v)\| - \|\overline{g}_a(x) - g_a(u)\| - \|\overline{g}_a(y) - g_a(v)\| \geq \|u - v\| - 2D\varepsilon \geq \|x - y\| - 2\varepsilon - 2D\varepsilon \geq \|x - y\| - \frac{3}{C}$. □

The next lemma follows by very similar considerations and we omit the proof.

**Lemma 3.4**

(i) *For $a \neq b$, the Euclidean distance of $\Sigma_a$ and $\Sigma_b$ is at least $\rho_X(a, b) - \frac{2}{C}$, and in particular, $\Sigma_a \cap \Sigma_b = \emptyset$.*

(ii) *For any $a, b \in X$ and $x \in S$, we have $\|\overline{g}_a(x) - \overline{g}_b(x)\| \leq 2D_{\max}\Delta$.*

**Proof of Proposition 3.2(ii).** As was announced above, we can now apply Corollary 2.2 to the maps $\overline{g}_a$ with domain and range rescaled by $\frac{1}{R}$ (using $\delta = 3/CR < \frac{1}{4}$ and $2D_{\max}\Delta/R \leq \frac{1}{4}$). Letting $U_a$ denote the unbounded component of $\mathbb{R}^d \setminus \Sigma_a$, we can number the points of $X$ as $a_1, \ldots, a_n$ so that for $i < j$ we have $U_{a_i} \supset U_{a_j}$.

For $i = 1, 2, \ldots, n-1$ we define $\delta_i$ as the (Euclidean) distance of $\Sigma_{a_i}$ from $\Sigma_{a_{i+1}}$, and we define a mapping $f: X \to \mathbb{R}^1$ by $f(a_i) = \sum_{j=1}^{i-1} \delta_j$.

Assuming that the original mapping $g$ has distortion at most $D$, we will prove that $f$ has distortion at most $1.1D$. First we show that $f$ contracts distances by a factor of at most $1.1$. Lemma 3.4(i) gives $\delta_i \geq \rho_X(a_i, a_{i+1}) - 2/C \geq \rho_X(a_i, a_{i+1})/1.1$ (assuming $C$ large). The triangle inequality then shows that $|f(a_i) - f(a_j)| \geq \rho_X(a_i, a_j)/1.1$ for all $i, j$.

Next, we want to bound the Lipschitz constant of $f$. Let us fix a point $v_0 \in V$ and let us abbreviate $x_i := g(a_i, v_0)$. Let us fix $i < j$ and let us consider the line segment $x_i x_j$. We note that whenever $k$ lies between $i$ and $j$, the segment $x_i x_j$ intersects $\Sigma_{a_k}$. This is because $\Sigma_{a_j} \subset U_{a_k}$, while $\Sigma_{a_i} \cap U_{a_k} = \emptyset$. Thus, for each $k$, $i \leq k \leq j$, we can fix a point $y_k \in \Sigma_{a_k}$ on $x_i x_j$, where $y_i = x_i$ and $y_j = x_j$ (we note that $y_k$ also depends on $i$ and $j$). Then

$$\begin{aligned}|f(a_j) - f(a_i)| &= \sum_{k=i}^{j-1} \delta_k \leq \sum_{k=i}^{j-1} \|y_{k+1} - y_k\| = \|x_i - x_j\| \\ &\leq D\rho_Y((a_i, v_0), (a_j, v_0)) = D\rho_X(a_i, a_j)\end{aligned}$$

since $g$ is $D$-Lipschitz.

It remains to show how $f$ can be found from $g$ in polynomial time. First we need to sort the $\Sigma_a$. To compare $\Sigma_a$ and $\Sigma_b$, we can compute a point with the minimum $x_1$-coordinate, say, of $\Sigma_a \cup \Sigma_b$ and see if it lies in $\Sigma_a$ or $\Sigma_b$ (here we can use a property which follows from the proof of the Kirszbraun theorem, namely, that we may assume $\overline{g}_a(S) \subseteq \text{conv}(g_a(V))$, which implies that the smallest point of $\Sigma_a$ lies in $g_a(V)$). Then we can approximate the distance of $\Sigma_a$ to $\Sigma_b$ by the distance of the finite sets $g_a(V)$ and $g_b(V)$; this causes a small additive error which can increase the distortion of $f$ only negligibly. This concludes the proof of Proposition 3.2. $\square$

## 4 Punctured pseudospheres

For the stronger inapproximability result for dimensions 3 and higher, Theorem 1.2, we will need a nesting property not only for images of dense sets in spheres, but also for images for dense sets in other shapes.

In this section we develop a version of the nesting lemma that covers all of our applications. The definitions are tailored to these applications. In order to reduce the number of parameters, we use the same bound $\varepsilon$ for several independent small quantities; if we were aiming at tighter bounds in the inapproximability results, we could fine-tune each of these quantities independently.

Let $S \subseteq \mathbb{R}^d \setminus \{0\}$ be a set, and let $\varepsilon > 0$. We call a set $V \subseteq S$ $\varepsilon$-*angularly dense* in $S$ if for every $x \in S$ there exists $v \in V$ with $\|x - v\| \leq \varepsilon \|v\|$.

We call a set $P \subseteq \mathbb{R}^d$ $\varepsilon$-*angularly small with respect to* a set $V \subset \mathbb{R}^d \setminus \{0\}$ if there is a choice of a radius $r_v \geq 0$ for every $v \in V$ such that $P \subseteq \bigcup_{v \in V} B(v, r_v)$ (where $B(x, r)$ denotes the ball of radius $r$ centered at $x$) and $\sum_{v \in V} \frac{r_v}{\|v\|} \leq \varepsilon$ (this is a wasteful definition; aiming at more precise quantitative results, we would take $(r_v/\|v\|)^{d-1}$ instead of $r_v/\|v\|$, for example).

For our purposes, a *pseudosphere* is a set $S \subset \mathbb{R}^d$ homeomorphic to an $S^{d-1}$ such that

- the hole of $S$ contains 0, and

- there is a retraction $r_S$ of $\mathbb{R}^d \setminus \{0\}$ onto $S$ (i.e., $r_S\colon \mathbb{R}^d \setminus \{0\} \to S$ is a continuous map whose restriction on $S$ is the identity map).

A *punctured pseudosphere* is a pair $(S, P)$, where $S$ is a pseudosphere and $P \subseteq S$, the "punctures" of the pseudosphere, is a subset of $S$, which we will assume to be small in a suitable sense.

**Proposition 4.1 (Nesting lemma for punctured pseudospheres)** *Let $d \geq 2$, let $D \geq 1$ and let $\varepsilon := \frac{1}{16D}$, let $(S, P)$ be a punctured pseudosphere in $\mathbb{R}^d$, let $V \subset S$ be an $\varepsilon$-angularly dense set in $S$, let us assume that $P \subseteq S$ is $\varepsilon$-angularly small w.r.t. $V$, that $P \cap V = \emptyset$, and that $S \setminus P$ is path-connected, and let $f_1, f_2, \ldots, f_n\colon S \to \mathbb{R}^d$ be maps such that:*

- Each $f_i$ is $D$-Lipschitz.

- Each $f_i$ restricted to $V$ is noncontracting.

- We have $\|f_i(v) - f_j(v)\| \leq \frac{1}{4}\|v\|$ for all $v \in V$ and all $i,j$.

- Setting $\Sigma_i := f_i(S)$ and $\Sigma_i^* := f_i(S \setminus P)$, we have $\Sigma_i \cap \Sigma_j^* = \emptyset$ for all $i \neq j$.

Let $U_i$ denote the unbounded component of $\mathbb{R}^d \setminus \Sigma_i$, and let us define a relation $\preceq$ on $[n]$ by setting $i \preceq j$ if either $i = j$ or $\Sigma_j^* \subset U_i$. Then $\preceq$ is a linear ordering on $[n]$.

Moreover, $\preceq$ is independent of the behavior of the $f_i$ on the punctures, in the following sense: If $\tilde{f}_1, \ldots, \tilde{f}_i$ are $D$-Lipschitz mappings $S \to \mathbb{R}^d$ such that $f_i(x) = \tilde{f}_i(x)$ for all $x \in S \setminus P$ and all $i$ (in particular, $\Sigma_i^* = f_i(S \setminus P) = \tilde{f}_i(S \setminus P)$), and $\Sigma_i^* \cap \tilde{f}_j(S) = \emptyset$ for all $i \neq j$, then the linear ordering induced by the $\tilde{f}_i$ is the same as $\preceq$.

**A basic example.** Since this proposition is rather technical, let us present a basic example of a setting in which it will be applied. Let $C$ be a long cylinder in $\mathbb{R}^d$ of a large radius $R$ (see Fig. 4 left), and let $V$ be a set that is $\varepsilon$-dense in the lateral surface of $C$. With this $V$ we make a construction similar to the one in the proof of Theorem 1.1 above. We set $Y = [n] \times V$, and we define a metric on $Y$ by $\rho_Y((i,v),(i',v')) = \|v - v'\| + \delta_{ii'}$, where $\delta_{ii'}$ is the Kronecker delta (equal to 0 for $i = i'$ and to 1 otherwise).

We assume $\varepsilon \ll 1 \ll R$, and so we expect that if $g: Y \to \mathbb{R}^d$ is a $D$-embedding with $D$ not too large, the images of the $n$ copies of $V$ in $Y$ have to look like "nested cylinders". Let $g_i: V \to \mathbb{R}^d$ be the slice of $g$ corresponding to $i$.

In order to speak of "inside and outside" of these images, we let $S$ to be the whole surface of the cylinder $C$, including the top and the bottom, and we extend each $g_i$ to a $D$-Lipschitz $\overline{g}_i: S \to \mathbb{R}^d$. Now the images of the lateral surface $L$ of the cylinder under the $\overline{g}_i$ are disjoint (with an appropriate setting of $r, \varepsilon, D$), but we don't have much control over the images of the top and bottom, as is schematically indicated in Fig. 4 on the right (in a 2-dimensional cross-section). However, if we define $P$ as a suitable neighborhood, of radius about $DR$, of the top and bottom of $C$, then it can be checked that $\overline{g}_i(S \setminus P)$ avoids $\overline{g}_j(S)$ for $i \neq j$. In this situation, if $C$ is sufficiently long, Proposition 4.1 allows us to conclude that the images of the $\overline{g}_i$ are nested (in the sense defined in the proposition).

**Towards the proof of Proposition 4.1.** Let us define

$$\Omega := \bigcup_{v \in V} B(f_1(v), \tfrac{1}{2}\|v\|).$$

As the next lemma shows, $\Omega$ is a neighborhood of all the $\Sigma_i$, and to some extent it will play the role of $\Omega_{\frac{1}{4}}$ from the proof of Theorem 2.1(i)–(ii).

**Lemma 4.2** *We have $\Sigma_i \subseteq \Omega$ for all $i$, and all the $f_i$ are homotopic as maps $S \to \Omega$.*

**Proof.** (A routine "$\varepsilon$-density" argument.) Given a point $f_i(x) \in \Sigma_i$, we choose $v \in V$ with $\|x - v\| \leq \varepsilon\|v\|$. Then $\|f_i(x) - f_1(v)\| \leq \|f_i(x) - f_i(v)\| + \|f_i(v) - f_1(v)\| \leq D\varepsilon\|v\| + \frac{1}{4}\|v\| < \frac{1}{2}\|v\|$, and hence $f_i(x) \in B(f_1(v), \frac{1}{2}\|v\|)$. Similarly $f_1(x) \in B(f_1(v), \frac{1}{2}\|v\|)$, so the segment $f_1(x)f_i(x)$ is contained in $\Omega$ for every $x$, and thus $f_i \sim f_1$ as claimed. □

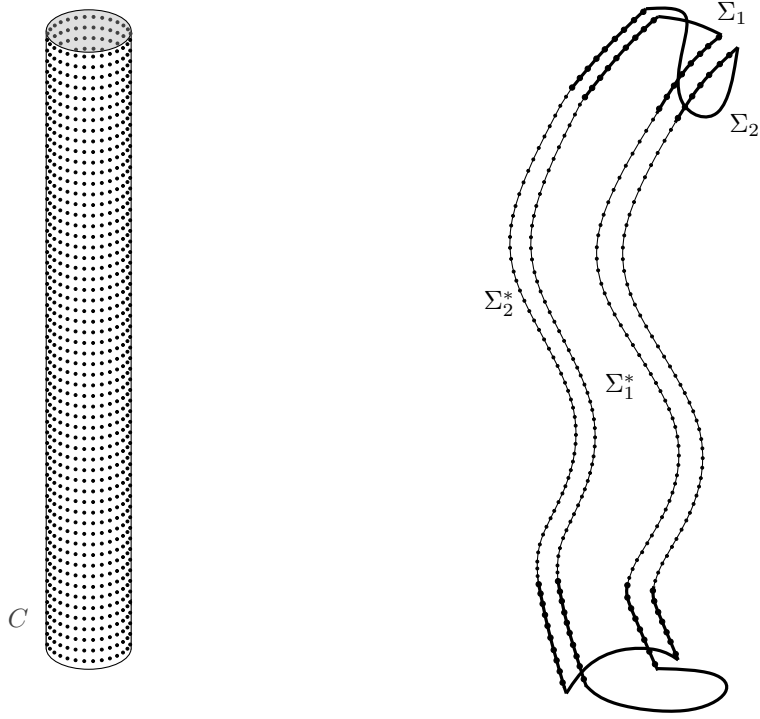

Figure 4: The set $V$ (left), and an embedding of $Y$ and its extension (right).

**Lemma 4.3** *There is a point $c \notin \Omega$ that lies in a hole $H_i$ of $\Sigma_i$ for each $i$.*

**Proof.** We imitate the proof of Theorem 2.1(ii). We set $N := f_1(V)$, for every $z \in N$ we fix $g(z) \in f_1^{-1}(z)$, and we fix a 1-Lipschitz extension $\overline{g} \colon \Omega \to \mathbb{R}^d$ of $g$.

We check that $0 \notin \operatorname{Im} \overline{g}$: Assuming $0 = \overline{g}(y)$, $y \in \Omega$, we choose $v \in V$ with $y \in B(f_1(v), \frac{1}{2}\|v\|)$. Then $\|v\| = \|\overline{g}(y) - \overline{g}(f_1(v))\| \leq \|y - f_1(v)\| \leq \frac{1}{2}\|v\|$, a contradiction since $v \neq 0$.

Next, we claim that $\overline{g} f_1 \sim \operatorname{id}_S$ as maps $S \to \mathbb{R}^d \setminus \{0\}$. It suffices to verify that the segment connecting $x$ to $\overline{g}(f_1(x))$ doesn't contain $0$, for all $x \in S$. As usual, we fix $v \in V$ with $\|x - v\| \leq \varepsilon \|v\|$; then we also have $\|v\| \leq \|x\|/(1-\varepsilon) \leq 2\|x\|$. We estimate $\|\overline{g}(f_1(x)) - x\| \leq \|\overline{g}(f_1(x)) - \overline{g}(f_1(v))\| + \|\overline{g}(f_1(v)) - x\| = \|\overline{g}(f_1(x)) - \overline{g}(f_1(v))\| + \|v - x\| \leq D\|v - x\| + \varepsilon\|v\| \leq (D+1)\varepsilon\|v\| \leq 4D\varepsilon\|x\| < \frac{1}{2}\|x\|$, and $\overline{g}f_1 \sim \operatorname{id}_S$ follows.

Next, we set $h := r_S \overline{g}$, where $r_S$ is the retraction of $\mathbb{R}^d \setminus \{0\}$ onto $S$. It follows that $hf_1 \sim \operatorname{id}_S$ as well (compose the homotopy witnessing $\overline{g}f_1 \sim \operatorname{id}_S$ with $r_S$). With such an $h$ at our disposal, an argument very similar to the proof of Theorem 2.1(ii) finishes the proof of the lemma. □

**Proof of Proposition 4.1.** Let $c$ be a point as in Lemma 4.3; for notational convenience, we will assume that $c = 0$. Let $H_i$ denote the hole of $\Sigma_i$ that contains $0$.

For a set $X \subseteq \mathbb{R}^d \setminus \{0\}$, let $\mu(X)$ denote the "spatial angle" of $X$ as seen from $0$, i.e., the fraction of $S^{d-1}$ occupied by the central projection of $X$. We claim that if $x \neq 0$ and $r \leq \frac{1}{2}\|x\|$, then we have $\mu(B(x, r)) \leq r/\|x\|$. Indeed, a simple projection argument shows that it is enough to deal with the case $d = 2$, and this case follows easily with a bit of trigonometry.

We claim that $\mu(f_i(P)) \leq \frac{1}{4}$ for all $i$. To check this, we use the assumption that the "puncture set" $P$ is $\varepsilon$-angularly small w.r.t. $V$: If $r_v$ are the radii as in the definition of "$\varepsilon$-angularly small", then $f_i(P) \subseteq \bigcup_{v \in V} B(f_i(v), Dr_v)$, and thus

$$\mu(f_i(P)) \leq \sum_{v \in V} \frac{Dr_v}{\|f_i(v)\|}.$$

Since $0 \notin \Omega$, the definition of $\Omega$ gives $\|f_1(v)\| > \frac{1}{2}\|v\|$, and so $\|f_i(v)\| \geq \|f_1(v)\| - \|f_1(v) - f_i(v)\| \geq \frac{1}{2}\|v\| - \frac{1}{4}\|v\| = \frac{1}{4}\|v\|$. Then $\mu(f_i(P)) \leq 4D \sum_{v \in V} r_v/\|v\| \leq 4D\varepsilon = \frac{1}{4}$.

Now we are ready to prove that $\preceq$ is a linear ordering. We first look at two of the indices, say $i = 1$ and $j = 2$, and show that either $1 \preceq 2$ or $2 \preceq 1$. Let $U_{12} \subseteq U_1 \cap U_2$ denote the unbounded component of $\mathbb{R}^d \setminus (\Sigma_1 \cup \Sigma_2)$, and let $H_{12} \subseteq H_1 \cap H_2$ be the hole of $\Sigma_1 \cup \Sigma_2$ containing $0$.

The boundary $\partial U_{12}$ is contained in $\Sigma_1 \cup \Sigma_2$, and we have $\mu(\partial U_{12}) = 1$. It follows that $\partial U$ cannot be covered by $f_1(P) \cup f_2(P)$, and thus at least one of $\Sigma_1^* = f_1(S \setminus P)$ and $\Sigma_2^* = f_2(S \setminus P)$ is incident to $\partial U_{12}$; let us suppose $x \in \Sigma_2^* \cap \partial U_{12}$, for example. Then, since $\Sigma_2^* \cap \Sigma_1 = \emptyset$, we have $x \in U_1$, and since $\Sigma_2^*$ is path-connected, we obtain $\Sigma_2^* \subset U_1$. Hence $1 \preceq 2$.

By a similar argument, looking at $\partial H_{12}$, we get that one of $\Sigma_1^* \subset H_2$ or $\Sigma_2^* \subset H_1$ holds, and this shows that $1 \preceq 2$ and $2 \preceq 1$ cannot hold simultaneously.

It remains to verify transitivity of $\preceq$. To this end, we consider three indices, say $1, 2, 3$, and we suppose $1 \preceq 2 \preceq 3$. This means that $\Sigma_2^* \subset U_1$ and $\Sigma_3^* \subset U_2$. If we had $3 \preceq 1$ as well, then $\Sigma_1^* \subset U_3$. So $\Sigma_1^* \cup \Sigma_2^* \cup \Sigma_3^* \subset U_1 \cup U_2 \cup U_3$. But a measure argument as above shows that at least one of $\Sigma_1^*$, $\Sigma_2^*$, $\Sigma_3^*$ must appear on the boundary of the hole $H_{123}$ of $\Sigma_1 \cup \Sigma_2 \cup \Sigma_3$ that contains $0$, but this is impossible since $H_{123} \subseteq H_1 \cap H_2 \cap H_3$, and the latter is disjoint from $U_1 \cup U_2 \cup U_3$. This shows that $\preceq$ is indeed a linear ordering on $[n]$.

It remains to verify the assertion that $\preceq$ doesn't depend on the behavior of the $f_i$ on the puncture set $P$. Thus, let the $\tilde{f}_i$ be as in the proposition. We can replace the $f_i$ by the $\tilde{f}_i$ gradually one by one (since the already proved part of the proposition applies to $f_1, f_2, \ldots, f_i, \tilde{f}_{i+1}, \ldots, \tilde{f}_n$ and shows that they induce some linear ordering), and thus it suffices to consider $f_1$, $f_2$, and $\tilde{f}_2$ and show that $1 \preceq 2$ iff $1 \preceq' 2$, where $\preceq'$ is the relation induced by $f_1$ and $\tilde{f}_2$.

First, if $1 \preceq 2$, then $\Sigma_2^*$ lies in the unbounded component of $\mathbb{R}^d \setminus \Sigma_1$, and hence $1 \preceq' 2$ by definition. Second, if $2 \preceq 1$, then as shown above, $\Sigma_2^*$ is contained in a hole of $\Sigma_1$, and this means $2 \preceq' 1$. The proposition is proved. □

## 5 Stronger inapproximability for dimension 3: a warm-up

In this section we present a simple reduction, which provides an inapproximability result weaker than Theorem 1.2: in that theorem, we claim the hardness of distinguishing between $O(1)$-embeddability and $n^{\mathrm{const}/d}$-embeddability, while here we show hardness of distinguishing between $n^\varepsilon$-embeddability ($\varepsilon > 0$ arbitrary but fixed) and $n^{\mathrm{const}/d}$-embeddability.

We will use an algorithmic problem called BETWEENNESS, which is NP-complete according to Opatrny [24] (the beautiful reduction of 3-SAT to this problem is also reproduced in [10]). An instance of BETWEENNESS is a set $T$ of triples of the form $(i, j, k)$, $i, j, k \in [n]$, and the problem is to decide whether $T$ is *consistent*, i.e., whether there exists a linear ordering $\preceq$ of $[n]$ for which $i$ is between $j$ and $k$ for every $(i, j, k) \in T$ (that is, either $j \preceq i \preceq k$ or $k \preceq i \preceq j$).

It will be more convenient to reduce to NON-BETWEENNESS, whose instance has the same form as for BETWEENNESS but the meaning of $(i, j, k)$ is now "$i$ must *not* be between $j$ and $k$".

Each constraint $(i, j, k)$ in BETWEENNESS can be equivalently replaced by the two constraints $(j, i, k)$ and $(k, i, j)$ in NON-BETWEENNESS, and so NON-BETWEENNESS is NP-complete as well.

**The reduction.** Let $d \geq 3$ be fixed. Given an instance $T$ of NON-BETWEENNESS for $n$ elements and a bound $D$ for distortion, we construct a metric space $\mathbb{Y} = (Y, \rho_Y)$, with $|Y| \leq (nD)^{O(d)}$, such that:

- If $T$ is consistent, then $\mathbb{Y}$ is $O(n)$-embeddable in $\mathbb{R}^d$.

- If $T$ is not consistent, then $\mathbb{Y}$ is not $D$-embeddable in $\mathbb{R}^d$.

Setting $D = n^C$ for a large constant $C$, we get that it is NP-hard to distinguish between $O(n)$-embeddability and $n^C$-embeddability of $\mathbb{Y}$ (and the size of $\mathbb{Y}$ is of order $n^{C_0 C d}$ for an absolute constant $C_0$).

We fix suitable parameters $\varepsilon \ll 1 \ll R$, with $\varepsilon$ sufficiently small and $R$ sufficiently large in terms of $n$ and $D$, and we let $S$ be a $(d-1)$-dimensional sphere of radius $R$. We let $V$ be an $\varepsilon$-dense set in $S$, and similar to the example following Proposition 4.1, we set $Y_0 := [n] \times V$ and $\rho_{Y_0}((i, v), (i', v')) = \|v - v'\| + \delta_{ii'}$. We will refer to the set $\{i\} \times V$ as the *ith layer*. Next, we will modify $(Y_0, \rho_{Y_0})$ to obtain $\mathbb{Y}$; this modification reflects the structure of $T$.

We choose $|T|$ distinct points on $V$, sufficiently far from one another, corresponding to the triples in $T$. We will call these points the *loci*.

Let $(i, j, k) \in T$ and let $v = v_{(i,j,k)} \in V$ be the corresponding locus. We modify the metric space $(Y_0, \rho_{Y_0})$ near $v$ as follows:

- We make a puncture of radius 1 in each of the layers except for the $i$th, $j$th, and $k$th. That is, we remove from $Y_0$ all points $(\ell, u)$ with $\ell \notin \{i, j, k\}$ and $\|u - v\| \leq 1$.

- We connect the $j$th and $k$th layers by a (discrete) path $\pi_{v,j,k}$ of length 1. Namely, we set $t = \lfloor 1/\varepsilon \rfloor$, we consider a graph-theoretic path on vertices $p_0, p_1, \ldots, p_t$ with edges of length $1/t$, and we glue this path to $Y_0$ by identifying $p_0$ with $(j, v)$ and $p_t$ with $(k, v)$ (while $p_1, \ldots, p_{t-1}$ are new points).

Having made this modification for every triple of $T$, we call the resulting metric space $\mathbb{Y} = (Y, \rho_Y)$.

**Embeddability for consistent instances.** We assume that $T$ is consistent and, for notational convenience, that a linear ordering obeying all the constraints is the natural ordering $\leq$. Then we embed $\mathbb{Y}$ in $\mathbb{R}^d$ as sketched in Fig. 5 (the picture is 2-dimensional although the actual embedding is in dimension 3 or higher). The $i$th layer is mapped isometrically to the sphere of radius $R + i$ centered at 0, and the connecting paths $\pi_{v,j,k}$ are embedded on straight segments that pass through the punctures in the other layers. Clearly, this embedding incurs distortion $O(n)$.

***D*-embeddability implies consistency.** Next, let $g$ be an arbitrary $D$-embedding of $\mathbb{Y}$ into $\mathbb{R}^d$, which we assume to be noncontracting. We let $g_i: V \to \mathbb{R}^d$ be the corresponding $D$-embedding of the $i$th layer, and let $\overline{g}_i$ be a $D$-Lipschitz extension of $g_i$ on $S$.

For applying Proposition 4.1, we need to define the "puncture set" $P$ in such a way that $\overline{g}_i(S \setminus P)$ is guaranteed to be disjoint from $\overline{g}_j(S)$, $j \neq i$. Since the punctures in the construction of $\mathbb{Y}$ have diameter 2 and each $\overline{g}_i$ is $D$-Lipschitz, it suffices to choose $P \subseteq S$ as the union balls of radius $3D$ centered at all the loci (and intersected with $S$).

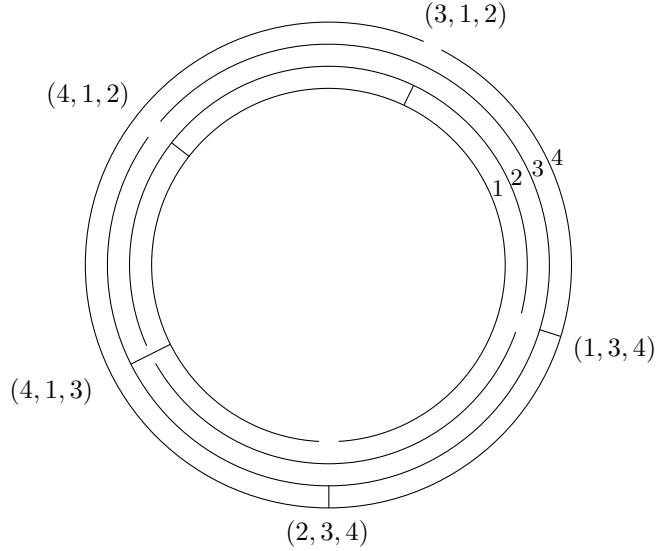

Figure 5: Embedding for a consistent instance ($n = 4$, $T = \{(3,1,2), (4,1,2), (4,1,3), (2,3,4), (1,3,4)\}$).

If $R$ is sufficiently large in terms of $D$, then the complement $S \setminus P$ is connected; this is *the* moment in the proof where we need the assumption $d \geq 3$ (since if we puncture 1-dimensional sphere more than once, it falls apart!). For metric reasons, we then have $\Sigma_i \cap \Sigma_j^* = \emptyset$ for $i \neq j$, where $\Sigma_i := \overline{g}_i(S)$, $\Sigma_i^* := \overline{g}_i(S \setminus P)$. It is straightforward to verify the assumptions of Proposition 4.1 with $f_i = \overline{g}_i$ (we note that $V$ here and in the proposition are the same, but the $\varepsilon$ in the proposition, equal to $\frac{1}{16D}$, need not be the same).

We thus conclude that the images $\overline{g}_i(S)$ are nested in the sense of Proposition 4.1, and we claim that the corresponding linear ordering $\preceq$ of $[n]$ is consistent for $T$. Indeed, for contradiction let us suppose that $(i,j,k) \in T$ but $j \preceq i \preceq k$, say. Then $\Sigma_k^*$ is contained in the unbounded component of $\Sigma_i$, while $\Sigma_j^*$ lies in a hole of $\Sigma_i$.

If $v$ is the locus of $(i,j,k)$, then none of the layers $i,j,k$ has a puncture near $v$, and thus for metric reasons, $\Sigma_j$, $\Sigma_i$, and $\Sigma_k$ do not intersect in the $3D$-neighborhood of $g_i(v)$. It follows that $g_j(v)$ lies in the same connected component of $\mathbb{R}^d \setminus \Sigma_i$ as $\Sigma_j^*$, i.e., in a hole, while $g_k(v)$ is in the unbounded component.

We now consider the image under $g$ of the discrete path $\pi_{v,j,k}$ in $\mathbb{Y}$, and we extend this image by segments to a continuous path $\gamma$ connecting $g_j(v)$ and $g_k(v)$. Then $\gamma$ intersects $\Sigma_i$ at a point $x$. On $\gamma$ we can find a point $g(p)$ at distance at most $D\varepsilon$ from $x$, where $p \in \pi_{v,j,k}$, and in $\Sigma_i$ there is a point of the form $g(i,u)$, $u \in V$, also at distance at most $D\varepsilon$ from $x$. But $\rho_Y((i,u),p) \geq 1$, and so $2D\varepsilon \geq 1$ since $g$ was assumed noncontracting—a contradiction for $\varepsilon$ sufficiently small. □

## 6 Proof of Theorem 1.2

Here we present a different reduction of NON-BETWEENNESS to approximate embeddability in $\mathbb{R}^d$, in which consistent instances yield $O(1)$-embeddability. The main idea is similar to the previous reduction: the linear ordering in NON-BETWEENNESS is encoded in nesting of suitable "discretized surfaces". The source of the $\Omega(n)$ distortion in the previous reduction was the nesting of all the surfaces at the same time.

Here we will allow simultaneous nesting of only at most 3 surfaces at a time. The surfaces won't be simply spheres, though, but rather each of them will resemble a network of branching pipes. We begin with a simple graph-theoretic lemma.

**Lemma 6.1** *For every natural number $n$ there is a graph $G$ of size polynomial in $n$ and subgraphs $G_1, G_2, \ldots, G_n$ of $G$ such that*

- *Each $G_i$, as well as each $G_i \cap G_j$, is a connected subgraph of $G$.*

- *No vertex of $G$ belongs to more than 3 of the $G_i$.*

- *For every unordered triple $\{i, j, k\}$, there is a vertex $a_{ijk} \in V(G_i) \cap V(G_j) \cap V(G_k)$.*

**Proof.** The vertex set of $G$ can be taken as $\binom{[n]}{2} \cup \binom{[n]}{3}$ (all 2-element and 3-element subsets of $[n]$), and the edges are of the form $\{\{i,j\}, \{i,j,k\}\}$, $i, j, k$ all distinct. We let $G_i$ be the subgraph induced by the set of all $S \in V(G)$ with $i \in S$. Verifying the required properties is immediate. □

**The construction.** Let $d \geq 3$ be fixed. Given an instance $T$ of NON-BETWEENNESS for $n$ elements and a parameter $D$ representing maximum distortion, we first construct an initial metric space $\mathbb{Y}_0$ that depends only on $n$ and $D$.

We choose parameters $\varepsilon \ll 1 \ll R_{\text{edge}} \ll R_{\text{vert}}$ (polynomially depending on $n$ and $D$, with the degree of the polynomial independent of $d$). We fix an embedding of the graph $G$ as in Lemma 6.1 into $\mathbb{R}^d$, where vertices are represented by points and edges by straight segments. We assume that the minimum edge length is sufficiently large compared to $R_{\text{vert}}$, the maximum edge length is bounded by $R_{\text{vert}}$ times a polynomial in $n$, the minimum distance of every two vertex-disjoint edges is much larger than $R_{\text{edge}}$, and that the minimum angle of two edges sharing a vertex is bounded below by an inverse polynomial in $n$.

We now "fatten" the embedded $G$: We replace each vertex $a \in V(G)$ by a ball $B_a$ of radius $R_{\text{vert}}$ and each edge $e$ by a cylinder $C_e$ of radius $R_{\text{edge}}$. We choose an $\varepsilon$-dense set $V$ in the boundary of the resulting solid (the union of all $B_a$ and all $C_e$). We let $V_a := V \cap \partial B_a$ and $V_e := V \cap \partial C_e$, and for $i \in [n]$

$$V_i := \left( \bigcup_{e \in E(G_i)} V_e \right) \cup \left( \bigcup_{a \in V(G_i)} V_a \right),$$

where the $G_i$ are the subgraphs as in Lemma 6.1. The metric space $\mathbb{Y}_0 = (Y_0, \rho_{Y_0})$ is given by

$$\begin{aligned} Y_0 &= \{(i, v) : i \in [n], v \in V_i\} \\ \rho_{Y_0}((i, v), (i', v')) &= \|v - v'\| + \delta_{ii'}. \end{aligned}$$

The $i$th layer of $Y_0$ is $\{i\} \times V_i$.

Now for every triple $(i, j, k) \in T$, we choose a point $v \in V_{a_{ijk}}$, not too close to any $V_e$, and we connect the points $(j, v)$ and $(k, v)$ by a discrete path of length 1 with spacing $\varepsilon$. (Since the vertices $a_{ijk}$ are indexed by *unordered* triples, while the triples in $T$ are ordered, we may need several such paths for a single vertex.) Adding such paths for all triples in $T$ yields the metric space $\mathbb{Y}$.

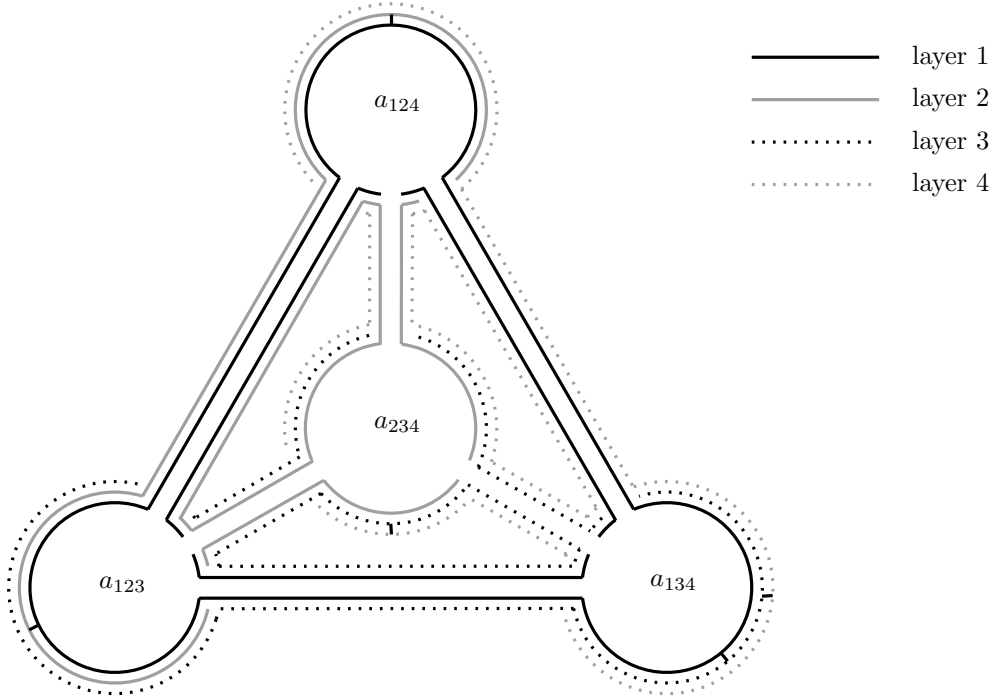

Figure 6: An embedding for a consistent instance.

**Consistent instances.** If $T$ is consistent, it is easy to embed $\mathbb{Y}$ in $\mathbb{R}^d$ with distortion $O(1)$. Rather than trying to formalize this, we refer to Fig. 6 for a (misleadingly planar) sketch for $n = 4$, with $T$ the same as in Fig. 5 (for $n = 4$, the graph $G$ can be taken very simple, as a $K_4$, with each $G_i$ a triangle).

**$D$-embeddability implies consistency.** We consider a noncontracting $D$-embedding $g\colon Y \to \mathbb{R}^d$. Using Proposition 4.1 as in Section 5, we get that for each vertex $a$ of $G$ the embedding defines a linear ordering of at most 3 layers present at $a$, and similarly for each edge of $G$.

Next, we check that these orderings are locally consistent between a vertex $a$ and the adjacent edges.

**Lemma 6.2** *Let $\preceq_a$ be the linear ordering of the layers present at a vertex $a$, and let $\preceq_e$ be the linear ordering of the layers at an edge $e$ incident to $a$. If layers $i$ and $j$ are present both at $a$ and at $e$, then $i \preceq_a j$ iff $i \preceq_e j$.*

**Proof.** For notational convenience we will consider $i = 1$ and $j = 2$.

We consider three punctured pseudospheres: $S_a := \partial B_a$, $S_e := \partial C_e$, $S_{ae} := \partial(B_a \cup C_e)$. The puncture sets $P_a, P_e, P_{ae}$ are defined as expected: $P_a \subset S_a$ is the union of suitable neighborhoods of the cylinders $C_{e'}$ for all edges $e'$ incident to $a$, $P_e \subset S_e$ consists of suitable neighborhoods of the top and bottom sides of the cylinder $C_e$, and $P_{ae} \subset S_{ae}$ is $P_a \cup P_e$ *minus* the parts of $P_e$ and $P_a$ surrounding the place where $C_e$ touches $B_a$; see Fig. 7.

We consider the restriction $g_i$ of the embedding $g\colon Y \to \mathbb{R}^d$ to the $i$th layer, $i = 1, 2$. We fix a $D$-Lipschitz map $\overline{g}_{i,a}\colon S_a \to \mathbb{R}^d$ that coincides with $g_i$ on $V_a$, and similarly for $\overline{g}_{i,e}$ and $\overline{g}_{i,ae}$;

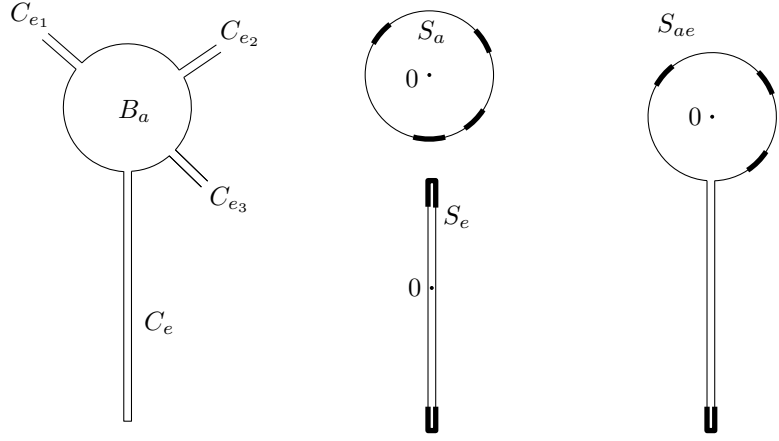

Figure 7: The ball $B_a$ and the cylinders adjacent to it (left); the punctured pseudospheres $S_a$, $S_e$, $S_{ae}$ (right); puncture sets indicated by a thick line.

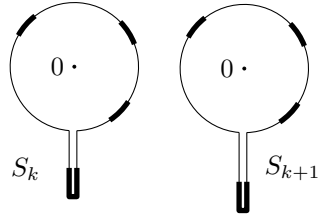

Figure 8: Two consecutive punctured pseudospheres $S_k$ and $S_{k+1}$.

moreover, we make sure that $\overline{g}_{i,a}$ and $\overline{g}_{i,e}$ coincide with $\overline{g}_{i,ae}$ on the shared parts of the domains (to this end, we can first extend $g_i$ to all of $S_a \cup S_e$ and then take $\overline{g}_{i,a}$, $\overline{g}_{i,e}$, and $\overline{g}_{i,ae}$ to be restrictions of this common extension).

The relation $\preceq_a$ is determined by the nesting of the images $\Sigma_{i,a} := \overline{g}_{i,a}(S_a)$, and similarly for $\preceq_e$. Instead of relating $\preceq_a$ and $\preceq_e$ directly, we relate both of them to $\preceq_{ae}$, which is the relation defined by the nesting of $\Sigma_{1,ae}$ and $\Sigma_{2,ae}$.

Let us first show that $\preceq_a$ and $\preceq_{ae}$ are the same. We will use the last part of Proposition 4.1, which tells us that the nesting relation doesn't depend on the behavior of the considered maps on the puncture set, and also the freedom to choose the puncture set.

We define a sequence of punctured pseudospheres $S_0 = S_a, S_1, \ldots, S_t = S_{ae}$ that interpolate between $S_a$ and $S_{ae}$: We start with the sphere $S_a$, and we gradually "grow" an attached cylinder from it, as in Fig. 8. The puncture sets $P_k$ are as indicated in the picture.

We let $\overline{g}_{i,k}\colon S_k \to \mathbb{R}^d$ be a suitable $D$-Lipschitz map that coincides with $\overline{g}_{i,ae}$ on the common part of their domain (where $\overline{g}_{i,0} = \overline{g}_{i,a}$ and $\overline{g}_{i,t} = \overline{g}_{i,ae}$). Let $\Sigma_{i,k} := \overline{g}_{i,k}(S_k)$.

For each $k = 0, 1, \ldots, t$, we thus get a "nesting" relation $\preceq_k$. If $\preceq_a$ and $\preceq_{ae}$ were not the same, we would get that for some $k$, $\preceq_k$ is different from $\preceq_{k+1}$.

To see that this is impossible, we choose a 1-Lipschitz map $h\colon S_{k+1} \to S_k$ that is the identity map on $S_{k+1} \cap S_k$ (we just contract a piece of the lateral surface of the cylinder $C_e$). Then we define $g_i' \colon S_{k+1} \to \mathbb{R}^d$ as the composition $\overline{g}_{i,k} h$.

On the one hand, if we use Proposition 4.1 with $S_{k+1}$ as the pseudosphere $S$, $P_{k+1}$ as the puncture set $P$, $\overline{g}_{i,k+1}$ as $f_i$, $i = 1, 2$, and $g_i'$ as $\tilde{f}_i$, $i = 1, 2$, we get that the nesting relation $\preceq'$ determined by $g_1'$ and $g_2'$ coincides with $\preceq_{k+1}$, since $\overline{g}_{i,k+1}$ agrees with $g_i'$ outside the puncture set.

On the other hand, the nesting relation depends solely on the images of the considered maps. In our case we have $g_i'(S_{k+1}) = \Sigma_{i,k} = \overline{g}_{i,k}(S_k)$, $i = 1, 2$. For the images of the "non-puncture sets" we have the inclusion $\Sigma_{i,k}^* = \overline{g}_{i,k}(S_k \setminus P_k) \subseteq g_i'(S_{k+1} \setminus P_{k+1})$, and the latter set is path-connected. Therefore, considering the way the nesting relation is defined in Proposition 4.1, we get that $\preceq'$ also coincides with $\preceq_k$. In conclusion, $\preceq_a$ is the same as $\preceq_{ae}$.

It remains to show that $\preceq_e$ and $\preceq_{ae}$ are also the same. We proceed similarly, but this time a single step is actually sufficient. We consider yet another punctured pseudosphere $S_{ea}$, which is $S_{ae}$ translated so that the origin is in the middle of the cylinder $C_e$ (as in $S_e$). The puncture set $P_{ea}$ of $S_{ea}$ includes the appropriately translated copy of $P_{ae}$ plus all the spherical part of $S_{ea}$ (we take advantage of the fact that the spherical part is angularly small in $S_{ea}$, provided that $C_e$ is sufficiently long compared to the radius of $S_a$).

The map $\overline{g}_{i,ea}$ is defined on $S_{ea}$ in the same way as $\overline{g}_{i,ae}$ on $S_{ae}$ (the only difference is the translation of the domain). The nesting relation $\preceq_{ea}$ obtained from these maps is the same as $\preceq_{ae}$ (since the images are the same, up to an enlargement of the puncture set).

The argument showing that $\preceq_{ea}$ coincides with $\preceq_e$ is then almost the same as the one above for $\preceq_k$ and $\preceq_{k+1}$, using a suitable 1-Lipschitz map $h' \colon S_{ea} \to S_e$ that collapses the spherical part of $S_{ea}$. This concludes the proof of the lemma. □

By Lemma 6.2, and since each $G_i \cap G_j$ is connected and nonempty, the local linear orderings at the vertices consistently define a relation $\preceq$ on $[n]$ such that for each $i \neq j$, exactly one of $i \preceq j$ and $j \preceq i$ holds. Since every three indices meet at a vertex of $G$, $\preceq$ is also transitive and thus a linear ordering. Finally, as in Section 5, the paths in $\mathbb{Y}$ make sure that $\preceq$ obeys all constraints in $T$.

It is not hard to see that $R_{\text{edge}}, R_{\text{vert}}, 1/\varepsilon$ can be bounded by a fixed polynomial in $n$ and $D$, of degree independent of $d$, and thus $|Y| = O((nD)^{Bd})$ for some universal constant $B$. This concludes the proof of Theorem 1.2. □

## 7 No Menger-type lemma for low-distortion planar embeddings

Here we prove Theorem 1.3. We set $w := c_1 \sqrt{\varepsilon}/k$ for a sufficiently small constant $c_1 > 0$. To construct $\mathbb{X} = (X, \rho_X)$, we consider a drawing of the graph $K_{3,3}$ (any other fixed nonplanar graph would do) as in the left part of Fig. 9 (here $w$ determines the width of the oval-shaped strip).

We place the $n$ points of $X$ on the edges as indicated, with a regular spacing of $\text{const}/n$. The distances under $\rho_X$ are given by the Euclidean distances of the points in the considered drawing, except for the points $p_1, \ldots, p_m, q_1, \ldots, q_m$ that fall in the dotted rectangle ($m$ is a fixed fraction of $n$). The distance of each $p_i$ or $q_i$ to a point outside the dotted rectangle is still the Euclidean distance in the drawing, but $\rho_X(p_i, q_j)$ is given by the Euclidean distances in the planar point configuration shown in Fig. 9 right, where the $p_i$ are placed on the top side of a rectangle of height $w$ and the $q_i$ on the bottom side.

Arguing as in [20], we can easily check that any embedding of $\mathbb{X}$ in the plane requires distortion $\Omega(nw) = \Omega(\sqrt{\varepsilon}n/k)$. Indeed, given a $D$-Lipschitz noncontracting mapping $f \colon X \to \mathbb{R}^2$, we extend it to a drawing of $K_{3,3}$ by connecting the images of the points of $X$ by straight segments. In every drawing of $K_{3,3}$, there are two vertex-disjoint edges $e, e'$ that cross. At distance at most $O(D/n)$ from the crossing we find images of two points $x, x' \in X$, $x$ on $e$ and $x'$ on $e'$. However, we have $\rho_X(x, x') \geq w$, and thus $D = \Omega(nw)$ as claimed.

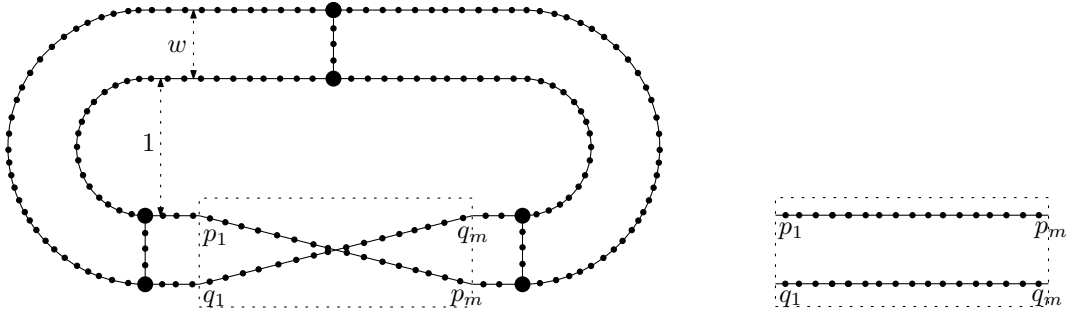

Figure 9: The construction for the proof of Theorem 1.3.

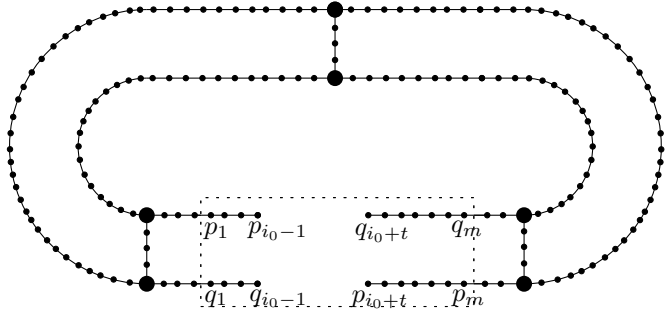

Figure 10: Embedding a $k$-point subspace in Theorem 1.3.

Next, let $\mathbb{Y}$ be a $k$-point subspace of $\mathbb{X}$. Then there is an $i_0$ such that none of $p_{i_0}$, $q_{i_0}$, $p_{i_0+1}$, $q_{i_0+1}, \ldots, p_{i_0+t-1}, q_{i_0+t-1}$ falls in $\mathbb{Y}$, where $t = \Omega(n/k)$. We embed $\mathbb{Y}$ as in Fig. 10. A simple calculation shows that the distortion of distances among the $p_i$ and $q_i$ is $1 + O((wn/t)^2) \leq 1 + \varepsilon$, while the distortion of distances involving some $p_i$ or $q_i$ and some other point is $O(1 + w) \leq 1 + \varepsilon$. □

**Remarks.** A similar example, with correspondingly weaker parameters, can be constructed for embeddings in $\mathbb{R}^d$ for every fixed $d \geq 2$, using the Van Kampen–Flores simplicial complexes nonembeddable in $\mathbb{R}^d$ (as in [20]) in place of $K_{3,3}$. The details are somewhat technical and there is no substantial new idea involved, so we prefer to omit this part.

Another natural question is, whether an "approximate Menger lemma" for embedding in the plane might hold for some interesting subclass of all metric spaces. Since the construction in the proof of Theorem 1.3 is based on a nonplanar graph, a natural class of interest are *planar-graph metrics*.[9]

For planar-graph metrics we have the following weaker analogy of Theorem 1.3: For all $n$,

---

[9] A metric space $(X, \rho_X)$ s a planar-graph metric if there is a planar graph $G = (V, E)$ with positive real weights on edges such that $X \subseteq V$ and $\rho_X$ is given by the shortest-path metric of $G$.

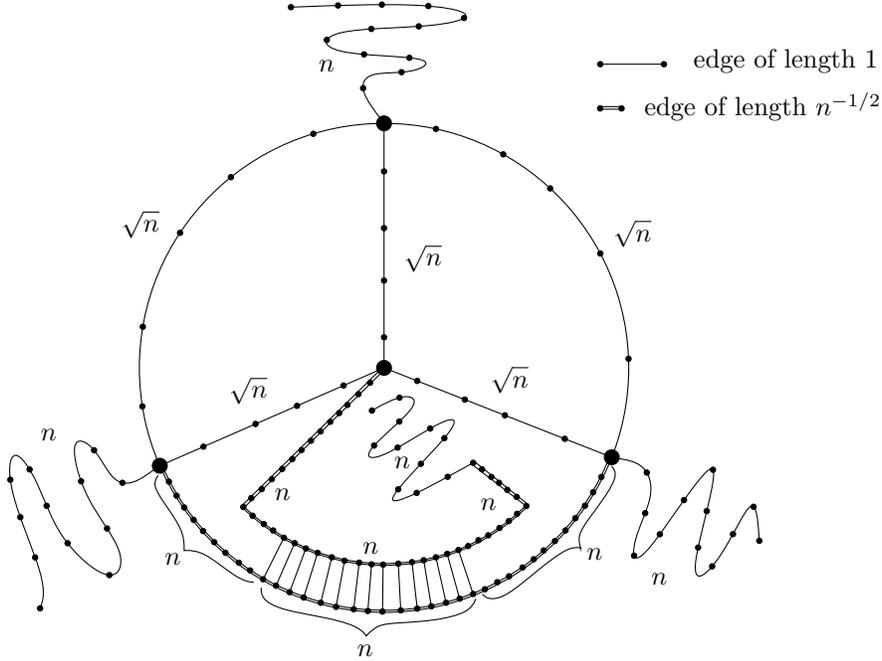

Figure 11: A construction for planar-graph metrics.

there are planar-graph metrics on $O(n)$ vertices whose embedding in $\mathbb{R}^2$ requires distortion $\Omega(\sqrt{n})$, while all subspaces on at most $\sqrt{n}$ points embed with distortion $O(1)$ (we don't get $1+\varepsilon$ as in Theorem 1.3, since planar-graph metrics usually don't embed almost isometrically in the Euclidean plane).

The construction is based on an idea from Bateni et al. [3] (in a simplified form); see Fig. 11. The metric space $\mathbb{X}$ is the vertex set of the depicted graph with the shortest-path metric. The edges drawn with a single line have length 1 and the edges drawn with a double line have length $n^{-1/2}$.

The reason for bad embeddability in the plane is as follows. The graph is essentially a subdivision of $K_4$ with a long path attached to each vertex. A low-distortion embedding of $\mathbb{X}$ in the plane yields a drawing of $K_4$ where vertex-disjoint edges don't cross, and in any such drawing of $K_4$ there is a vertex that is not incident to the outer face. Then the path attached to such a vertex doesn't have enough room in the inner face. The details are similar to an argument in [3] and we omit them. It is also easy to see that if we consider a subspace of $\sqrt{n}$ points, then there is a gap in the "ladder" in the bottom part of the graph of length $\Omega(\sqrt{n})$, and using such a gap, the path attached to the central vertex can cross to the outer face, which yields an $O(1)$-embedding.

## Acknowledgment

We thank Jussi Väisälä for providing us with a much better proof of Theorem 2.1(i) and (iii). We would like to thank Marianna Csörnyei for kindly answering a question of J. M. and suggesting a beautiful proof of the case $\delta = 0$ of Theorem 2.1(i).